\documentclass[aps,twocolumn,letter,showpacs,preprintnumbers,amsmath,amssymb,superscriptaddress,bibnotes]{revtex4-1}

\usepackage{graphicx,epsfig}
\usepackage{dcolumn}
\usepackage{bm}
\usepackage{textcomp}
\usepackage{amsmath}
\usepackage{amssymb}
\usepackage{color}
\usepackage{soul}

\begin{document}

\title{Weak antilocalization in quasi-two-dimensional electronic states of epitaxial LuSb thin films}

\author{Shouvik Chatterjee}
\email[Authors to whom correspondence should be addressed: ]{schatterjee@ucsb.edu, cjpalm@ucsb.edu}
\affiliation{Department of Electrical $\&$ Computer Engineering, University of California, Santa Barbara, CA 93106, USA}

\author{Shoaib Khalid}
\affiliation{Department of Physics and Astronomy, University of Delaware, Newark, DE 19716, USA}
\affiliation{Department of Materials Science and Engineering, University of Delaware, Newark, DE 19716, USA}
\author{Hadass S. Inbar}
\affiliation{Materials Department, University of California, Santa Barbara, CA 93106, USA}
\author{Aranya Goswami}
\affiliation{Department of Electrical $\&$ Computer Engineering, University of California, Santa Barbara, CA 93106, USA}
\author{Felipe Crasto de Lima}
\affiliation{Department of Materials Science and Engineering, University of Delaware, Newark, DE 19716, USA}
\author{Abhishek Sharan}
\affiliation{Department of Physics and Astronomy, University of Delaware, Newark, DE 19716, USA}
\affiliation{Department of Materials Science and Engineering, University of Delaware, Newark, DE 19716, USA}
\author{Fernando P. Sabino}
\affiliation{Department of Materials Science and Engineering, University of Delaware, Newark, DE 19716, USA}
\author{Tobias L. Brown-Heft}
\affiliation{Materials Department, University of California, Santa Barbara, CA 93106, USA}
\author{Yu-Hao Chang}
\affiliation{Materials Department, University of California, Santa Barbara, CA 93106, USA}

\author{Alexei V. Fedorov}
\affiliation{Advanced Light Source, Lawrence Berkeley National Laboratory, Berkeley, California 94720, USA}

\author{Dan Read}
\affiliation{School of Physics and Astronomy, Cardiff University, Cardiff CF24 3AA, UK}

\author{Anderson Janotti}
\affiliation{Department of Materials Science and Engineering, University of Delaware, Newark, DE 19716, USA}

\author{Christopher J. Palmstr\o m}
\email[Authors to whom correspondence should be addressed: ]{schatterjee@ucsb.edu, cjpalm@ucsb.edu}
\affiliation{Department of Electrical $\&$ Computer Engineering, University of California, Santa Barbara, CA 93106, USA}
\affiliation{Materials Department, University of California, Santa Barbara, CA 93106, USA}

\date{\today}

\begin{abstract}
Observation of large nonsaturating magnetoresistance in rare-earth monopnictides has raised enormous interest in understanding the role of its electronic structure. Here, by a combination of molecular-beam epitaxy, low-temperature transport, angle-resolved photoemission spectroscopy, and hybrid density functional theory we have unveiled the band structure of LuSb, where electron-hole compensation is identified as a mechanism responsible for large magnetoresistance in this topologically trivial compound. In contrast to bulk single crystal analogues, quasi-two-dimensional behavior is observed in our thin films for both electron and holelike carriers, indicative of dimensional confinement of the electronic states. Introduction of defects through growth parameter tuning results in the appearance of quantum interference effects at low temperatures, which has allowed us to identify the dominant inelastic scattering processes and elucidate the role of spin-orbit coupling. Our findings open up possibilities of band structure engineering and control of transport properties in rare-earth monopnictides via epitaxial synthesis.
\end{abstract}
\pacs{74.20.Rp, 74.25.Jb, 79.60.-i}
\maketitle
\section{Introduction}
Rare-earth monopnictides are of immense technological and scientific interest due to their potential applications in terahertz sources\cite{Bjarnason:2004jt,Sukhotin:2003kh,Salas:2017ia}, thermoelectrics\cite{Liu:2011hc}, solar cells\cite{Zide:2006gy}, plasmonics\cite{Krivoy:2018el}, and as buried epitaxial contacts\cite{Hanson:2006ir} when incorporated in III-V semiconductor heterostructures. Recently, it has been realized that these compounds also exhibit remarkably large magnetoresistance that has been attributed to either electron-hole compensation\cite{Zeng:2016et,Han:2017dw,Yang:2017fj}  or the presence of topologically nontrivial surface states\cite{FallahTafti:2016kl,Tafti:2015kj,Nayak:2016if,Singha:2017fi}. To elucidate the origins of these novel properties and for possible device applications it is thus imperative to gain an understanding of the electronic structure and scattering processes and how they are possibly modified in thin film geometries and/or nanostructures. 

Here, we present a demonstration of epitaxial synthesis of LuSb thin films using GaSb buffer layers on GaSb (001) substrates. We have observed Shubnikov de-Haas oscillations\cite{Shoenberg} from both electron and holelike carriers in our high mobility LuSb thin films, which are found to be in excellent correspondence with angle-resolved photoemission spectroscopy (ARPES) measurements and density functional theory (DFT) calculations with the Heyd-Scuseria-Ernzerhof (HSE06) hybrid functional\cite{Heyd:2003eg,HSE}. This has allowed us to experimentally map out the entire Fermi surface and determine Fermi wave vector ($k_{F}$), effective mass ($m^{*}$), and carrier concentration ($n$) of each of the bands constituting the Fermi surface. Observation of approximately equal concentration of electron and holelike carriers in LuSb coupled with the absence of any topological surface state in our ARPES measurements leads us to identify electron-hole compensation as the likely mechanism for large nonsaturating magnetoresistance observed in this compound. However, in contrast to bulk single crystals, quasi-two-dimensional behavior is observed for all the electronic bands in epitaxial films even with thicknesses as large as $\approx$ 14 nm. This is further corroborated by the observation of two-dimensional weak antilocalization (WAL) effects at low temperatures that also underscores the importance of spin-orbit scattering in this compound. Phase coherence length was found to be limited by electron-phonon scattering down to 2K. However, in comparison to bulk single crystals, Debye temperature ($\Theta_{D}$) is substantially reduced in our thin films. At low temperatures, characteristic phonon wavelength is found to be larger than the film thickness, placing phonons also in the 2D limit.

\section{Synthesis and Sample Characterization}

\begin{figure}
\includegraphics[width=0.9\columnwidth]{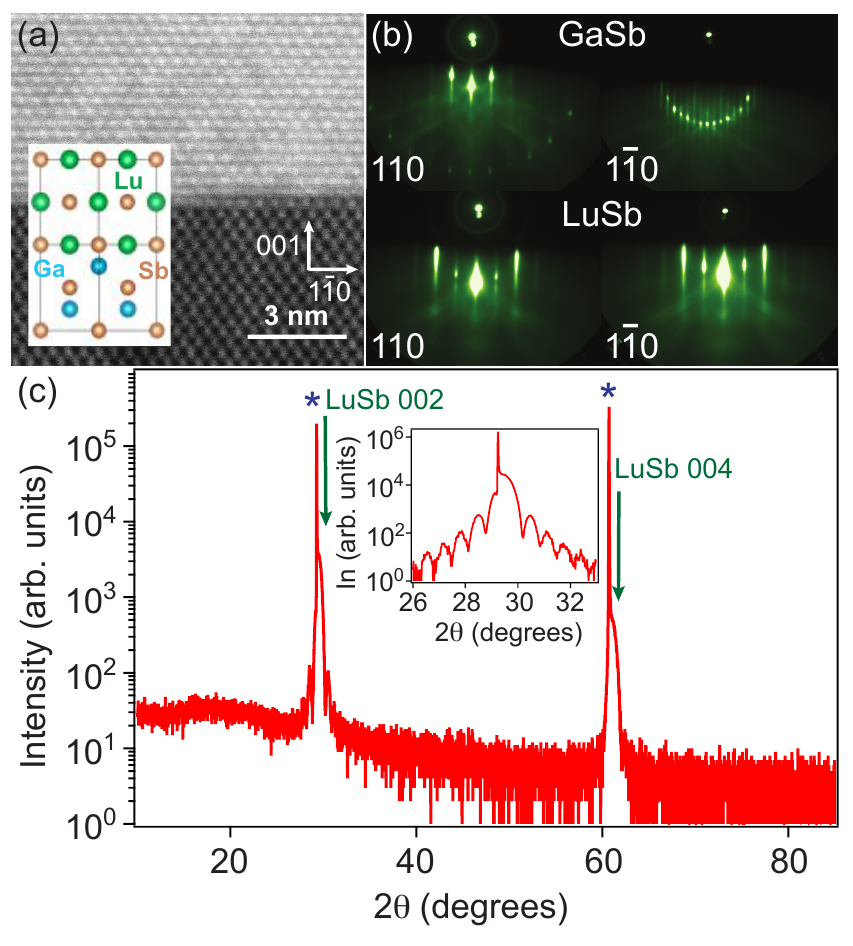}
\caption{Structural characterization of LuSb/GaSb (001) thin films. (a) HAADF-STEM image along the [110] zone axis. Inset shows the schematic of a proposed atomic arrangement across the GaSb-LuSb interface when viewed along the [110] direction. (b) RHEED images recorded after completion of growths of GaSb and LuSb epitaxial layers both along the [110] and [1$\bar{1}$0] azimuths. (c) Out-of-plane $\theta$-2$\theta$ XRD scan establish that our thin film is single phase. Substrate peaks are marked by asterisks. Inset shows thickness fringes around the (002) LuSb Bragg peak.}
\label{fig:Growth}
\end{figure}

Epitaxial thin films of LuSb (6.055 \AA) were synthesized on GaSb (6.095 \AA) substrates that are nearly perfectly lattice matched. In addition, the surface atomic arrangement of the antimony (Sb) atoms on the (001) surface provides an excellent template for epitaxial integration of LuSb atomic layers resulting in a continuous Sb sublattice. High angle annular dark field scanning transmission electron microscopy (HAADF-STEM) image shown in Fig.~\ref{fig:Growth}(a) confirms excellent quality of our thin films with minimal interdiffusion at the interface. By changing the substrate temperature and/or Lu:Sb beam flux ratio during growth we are able to introduce defects in our thin films that result in the appearance of quantum interference effects at low temperatures\cite{suppl}. RHEED images from such thin film samples are  shown in Fig. \ref{fig:Growth}(b). Sb-rich c(2$\times$6) reconstruction is observed for GaSb buffer layers that quickly changes to a (1$\times$1) reconstruction expected from stoichiometric rock-salt LuSb atomic layers. Out-of-plane $\theta$ - 2$\theta$ x-ray diffraction (XRD) further confirms that these thin films are single phase with (001) LuSb out-of-plane orientation. Smooth surfaces of our thin films results in thickness fringes from which we estimate a film thickness of 14.2 nm. All transport measurements presented in this work are from samples that exhibited quantum interference effects at low temperatures unless mentioned otherwise.

\section{Transport properties and electronic structure}

\begin{figure}
\includegraphics[width=0.9\columnwidth]{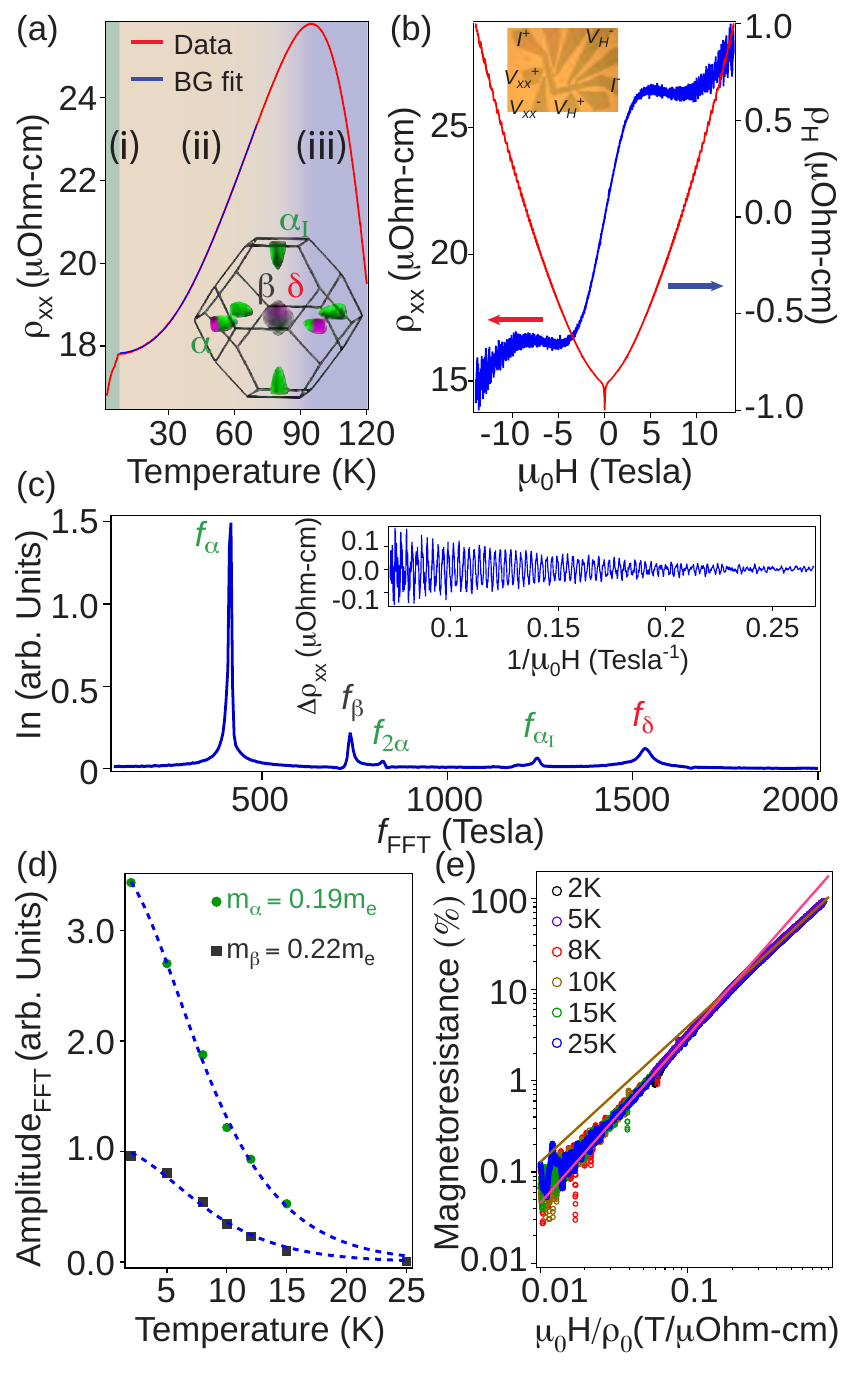}
\caption{Transport and fermiology in LuSb/GaSb (001) thin films. (a) Longitudinal resistivity as a function of temperature indicating three different regimes: (i) substrate dominated (blue background), (ii) LuSb dominated (brown background), (iii) LuSb dominated, where quantum corrections become significant (green background). Blue line is a fit to the data in region (ii), between 8K and 70K, to the Bloch-Gr{\"u}neissen functional form. Inset shows the calculated Fermi surface of LuSb. Current is applied along the [110] crystallographic direction. (b) Longitudinal and Hall resistivity as a function of magnetic field perpendicular to the sample plane. Optical micrograph of a hall bar device is shown in the inset. (c) Fast Fourier transform (FFT) of the quantum oscillation reveals three distinct frequencies corresponding to $\alpha$, $\beta$, and $\delta$ Fermi surface pockets. (d) Temperature dependence of the amplitudes of two main peaks in the FFT spectra in (c). Blue dotted lines are fits to thermal damping of the oscillations, as described in the main text. (e) Kohler's plot for magnetoresistance curves at different sample temperatures. Red and brown lines are linear fits to the data for low field, high temperature and high field, low temperature regimes, respectively.}
\label{fig:QO}
\end{figure}

WAL effect becomes manifest in our transport measurements below 8K leading to a dramatic drop in resistivity [Fig. \ref{fig:QO}(a)] that is readily suppressed on application of magnetic field perpendicular to the sample plane [Fig. \ref{fig:QO}(b)]. Temperature dependence of resistivity follows Bloch Gr{\"u}neissen functional form\cite{Ziman} with estimated Debye temperature of $\Theta_{D}$ = 267 K, which is much smaller in magnitude compared to what has been found in single crystals\cite{Pavlosiuk:2017fh}, and in other Lu monopnictides\cite{ogaA:2014km}. Transverse magnetoresistance shows nonsaturating behavior and reaches 110$\%$ at 14 T field. Magnetoresistance curves taken at different temperatures follow Kohler's scaling\cite{Ziman} $(\rho(B) - \rho(0))/\rho_{0}$ = $c(B/\rho_{0})^{m}$, shown in Fig. \ref{fig:QO}(e), indicating single dominant scattering process. However, unlike in single crystals the value of the exponent $m$ changes from 1.835 in the low field, high temperature regime to 1.486 in the high field, low temperature regime. Hall resistance shows multicarrier behavior as is expected from a compensated semimetal such as LuSb. Longitudinal and Hall resistivities were used to estimate mobility ($\mu$) and diffusion coefficient (\textit{D}) of each of the electronic bands (see Table I and Ref. \cite{suppl}).

\begin{table*}
  \caption{Transport parameters in LuSb}
  \label{tbl:example}
  \begin{ruledtabular}
  \begin{tabular}{cccccccc}
    \footnote{FS denotes Fermi surface, n$_{SdH}$ and n$_{DFT}$ are the carrier concentrations obtained from SdH and DFT, respectively, v$_{F}$ is the Fermi velocity, $\mu$ is the mobility, \textit{D} is the diffusion constant, $l_{e}$ is the elastic scattering length, and $B_{e}$ is the corresponding characteristic magnetic field. v$_{F}$, $\mu$, \textit{D}, $l_{e}$, and $B_{e}$ are calculated from data taken at 8 K \cite{suppl}. Average value of $B_{e}$ is calculated as $B_{e,avg} = \frac{\hbar}{4eD_{avg}\tau_{avg}}$, where the average is taken over all of the electronic bands in LuSb.}FS & n$_{SdH}$(cm$^{-3}$) & n$_{DFT}$(cm$^{-3}$)  & v$_{F}$(m/s) & $\mu$(cm$^{2}$/Vs) & \textit{D}(m$^{2}$/s)  & $l_{e}$(nm) & $B_{e}$(T) \\
    \hline
    $\alpha$  & 1.435$\times$10$^{20}$ & 1.45$\times$10$^{20}$ & 1.01$\times$10$^{6}$ & 5.42$\times$10$^{2}$ & 0.06 & 59 & 0.047   \\
    $\beta$ & 1.15$\times$10$^{20}$ & 1.17$\times$10$^{20}$ & 7.893$\times$10$^{5}$ & 4.078$\times$10$^{3}$ & 0.318 & 403 & 0.001 \\
    $\delta$  & 3.43$\times$10$^{20}$ & 2.92$\times$10$^{20}$ & 6.367$\times$10$^{5}$ & 2.036$\times$10$^{3}$ & 0.188 & 295  &  0.002\\
 
  \end{tabular}
\end{ruledtabular}
\end{table*}

We observe clear evidence of Shubnikov de-Haas (SdH) oscillations for fields stronger than 3.5 T, which is extracted after subtracting a smooth fifth order polynomial from the magnetoresistance data, shown in the inset of Fig. 2(b). Characteristic frequencies corresponding to an electron pocket at the zone edge ($\alpha$, $\alpha_{I}$) and two hole pockets at the zone center ($\beta$, $\delta$) are identified that match very well with both ARPES measurements and predictions from DFT calculations, summarized in Table II. Our results indicate that LuSb is a compensated semimetal with $n_{holes}/n_{electrons}$ = 1.06. From thermal damping of the amplitude of SdH oscillations\cite{Shoenberg} $R(T) = \frac{\lambda m^{*}T/B}{sinh({\lambda m^{*}T/B})}$, where $\lambda = \frac{2\pi^{2}k_{B}m_{e}}{e\hbar}$, we estimate the effective masses for the $\alpha$ and $\beta$ pocket to be 0.19m$_{e}$ and 0.22m$_{e}$, respectively.

\begin{figure*}
\includegraphics[width=0.9\textwidth]{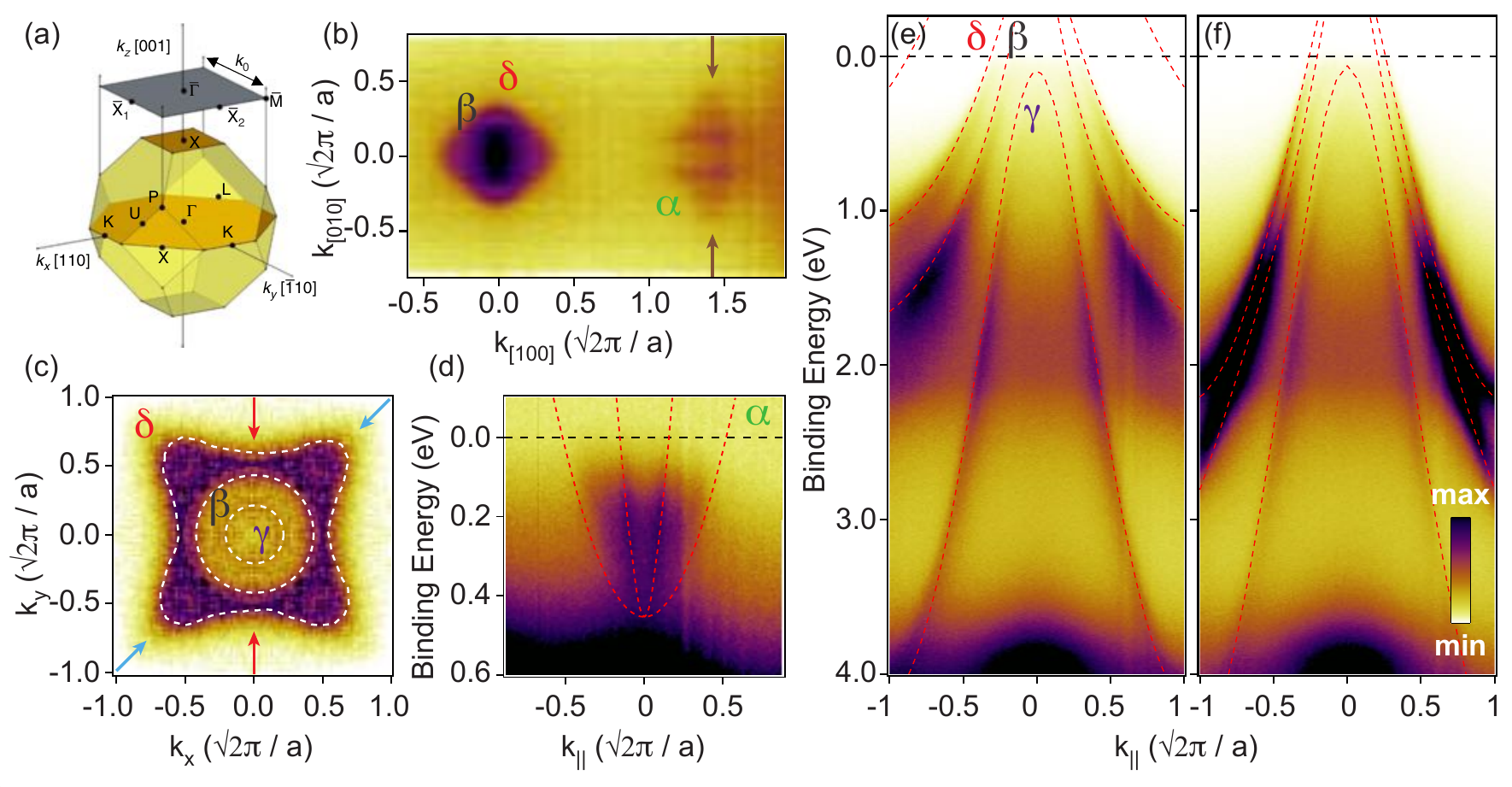}
\caption{ARPES spectra of LuSb/GaSb (001) thin films. (a) Bulk three-dimensional Brillouin zone of LuSb and its surface projection showing high-symmetry points. (b) Two-dimensional Fermi surface map near the bulk $\Gamma$ point showing both holelike($\beta$,$\delta$) and electronlike($\alpha$) Fermi surface sheets. (c) Two-dimensional map near the bulk $\Gamma$ point at a binding energy of 0.495eV illustrating anisotropy of the $\delta$ pocket. (d) \textit{E - k} spectral map along $\mathbf{\bar{\Gamma}}$-$\mathbf{\bar{M}}$-$\mathbf{\bar{\Gamma}}$ as indicated by brown arrows in panel (b). \textit{E - k} spectral maps along (e) $\mathbf{\bar{M}}$-$\mathbf{\bar{\Gamma}}$-$\mathbf{\bar{M}}$ and (f) $\mathbf{\bar{X}}$-$\mathbf{\bar{\Gamma}}$-$\mathbf{\bar{X}}$ indicated by blue and red arrows in (c), respectively. Red dotted lines are calculated band dispersions from DFT.}
\label{fig:ARPES}
\end{figure*}

In Fig. \ref{fig:ARPES}, we present \textit{E - k} spectral map along both $\mathit{\mathbf{\bar{M}}-\mathbf{\bar{\Gamma}}-\mathbf{\bar{M}}}$ and $\mathit{\mathbf{\bar{X}}-\mathbf{\bar{\Gamma}}-\mathbf{\bar{X}}}$ directions of the surface Brillouin zone for the hole pockets and along $\mathit{\mathbf{\bar{\Gamma}}-\mathbf{\bar{M}}-\mathbf{\bar{\Gamma}}}$ for the electron pocket. Effective masses are determined from parabolic fittings of the band dispersions at the Fermi level\cite{suppl}, which are in agreement with SdH and DFT results. Surface projection of the elliptical electron pocket allowed us to estimate effective masses along both the semimajor and semiminor axes of the ellipse (see Table II). We observe three holelike bands near the $\bar{\Gamma}$ point with the third band ($\gamma$) completely below the chemical potential in agreement with our DFT calculations. We must highlight the importance of using hybrid functionals\cite{Heyd:2003eg} in DFT calculations of rare-earth monopnictides. Use of generalized gradient approximation (GGA)\cite{Perdew:1996iq} erroneously predicts that all three holelike bands in LuSb cross the Fermi level, in clear disagreement with both our ARPES and quantum oscillation results\cite{suppl}.

\begin{table*}
\begin{center}
\caption{Fermi Surface of LuSb}
\begin{ruledtabular}
\begin{tabular}{c@{\quad}ccc@{\quad}ccc}
 &\multicolumn{3}{c}{k$_{F}$ (\AA$^{-1}$)} & \multicolumn{3}{c}{m$^{*}$} \\
 FS\footnote{FS denotes Fermi Surface} & SdH & ARPES & DFT & SdH & ARPES & DFT\\
\hline
  $\alpha$ & 0.11($a$), 0.34($b$)\footnote{$a$ and $b$ indicates directions along the semiminor and semimajor axes of the elliptical $\alpha$ pocket, respectively.} & 0.1($a$), 0.38($b$) & 0.11($a$), 0.37($b$) & 0.19 & 0.09($a$), 1.02($b$) &  0.11($a$), 1.16($b$) \\
  $\beta$ & 0.15 & 0.12(\textit{1}), 0.12($\mathit{\bar{1}}$)\footnote{\textit{1} and $\mathit{\bar{1}}$ indicates [100] and [110] crystallographic directions, respectively.} & 0.15(\textit{1}), 0.15($\mathit{\bar{1}}$)  & 0.22 & 0.26(1), 0.26($\mathit{\bar{1}}$) & 0.23(1), 0.21($\mathit{\bar{1}}$) \\
  $\delta$ & 0.22 & 0.21(\textit{1}), 0.19($\mathit{\bar{1}}$) & 0.24(\textit{1}), 0.19($\mathit{\bar{1}}$) & \textemdash & 0.45(\textit{1}), 0.36($\mathit{\bar{1}}$) & 0.54(\textit{1}), 0.31($\mathit{\bar{1}}$)\\

\end{tabular}
\end{ruledtabular}
\end{center}
\end{table*}

Having established the fermiology of our LuSb thin films we now turn to the magnetotransport results. Angle-dependent magnetotransport shows quasi-two-dimensional behavior for all the electronic bands in LuSb in marked contrast to its bulk single crystal analogues\cite{Pavlosiuk:2017fh}. The angular dependence of the SdH frequencies follows a 2D Fermi surface model $f_{\theta} = f_{0} / cos\theta$, where $\theta$ is the angle between the magnetic field vector and normal to the sample plane and $f_{0}$ is the SdH frequency at $\theta$ = 0. Although a similar angle-dependent behavior is expected from the bulk elliptical $\alpha$ band\cite{Han:2017dw}, an ellipticity ($k_{semimajor}$ / $k_{semiminor}$) much greater than the measured value of $\approx$ 3 is required to satisfactorily fit the observed angular dependence. Observed onset fields at which SdH oscillations begin to appear for different angular orientation further lends support to its quasi-2D behavior\cite{suppl,Allen:1991ut}.\\

\section{Analysis of quantum interference effects}

Thin films can be treated as quasi-two-dimensional if the film thickness is smaller than the relevant length scales. Electronic mean free paths ($l_{e}$ ) for all the electronic bands are found to be greater than the film thickness (see Table I), placing classical diffusive transport in our films in the 2D limit. WAL effects appearing in our thin films can also be considered as two dimensional as the associated characteristic length scale is the phase coherence length ($l_{\phi}$), which is required to be much greater than $l_{e}$ for such effects to appear in the first place and hence, must also be greater than the thickness of our thin film.

Presence of quantum interference effects, such as WAL in two dimensions, leads to an additional contribution to low temperature electron conductance $\Delta G = Aln(T/T_{0}$)\cite{Hikami:wh}, $T_{0}$ being the characteristic temperature. The prefactor A is negative for strong spin-orbit scattering ($\tau^{-1}_{SO}$ $\gg$ $\tau^{-1}_{\phi}$, $\tau_{SO}$ and $\tau_{\phi}$  are the spin-orbit and dephasing time, respectively)(WAL), in agreement with our experimental observation, shown in Fig. 4(b). 
Next, we utilize temperature dependence of the quantum interference effects under a perpendicular magnetic field to estimate phase coherence lengths using Hikami-Larkin-Nagaoka (HLN) theory \cite{Hikami:wh} for two-dimensional electron gas in the diffusive limit  that assumes Elliot-Yafet (E-Y)\cite{Lee:1985,AA:1983,Elliott:1954fv,Yafet:1963} spin-orbit scattering mechanism.  The centrosymmetric rock-salt crystal structure of LuSb coupled with the lack of evidence of Rashba-split states in our ARPES data guarantees that both the Dresselhaus and the Rashba effects\cite{Winkler} are unimportant in LuSb precluding us from considering Dyakonov-Perel (D-P)\cite{Dynakov:1972} scattering mechanism as a likely origin for the observed WAL effects.

\begin{figure*}
\includegraphics[width=0.8\textwidth]{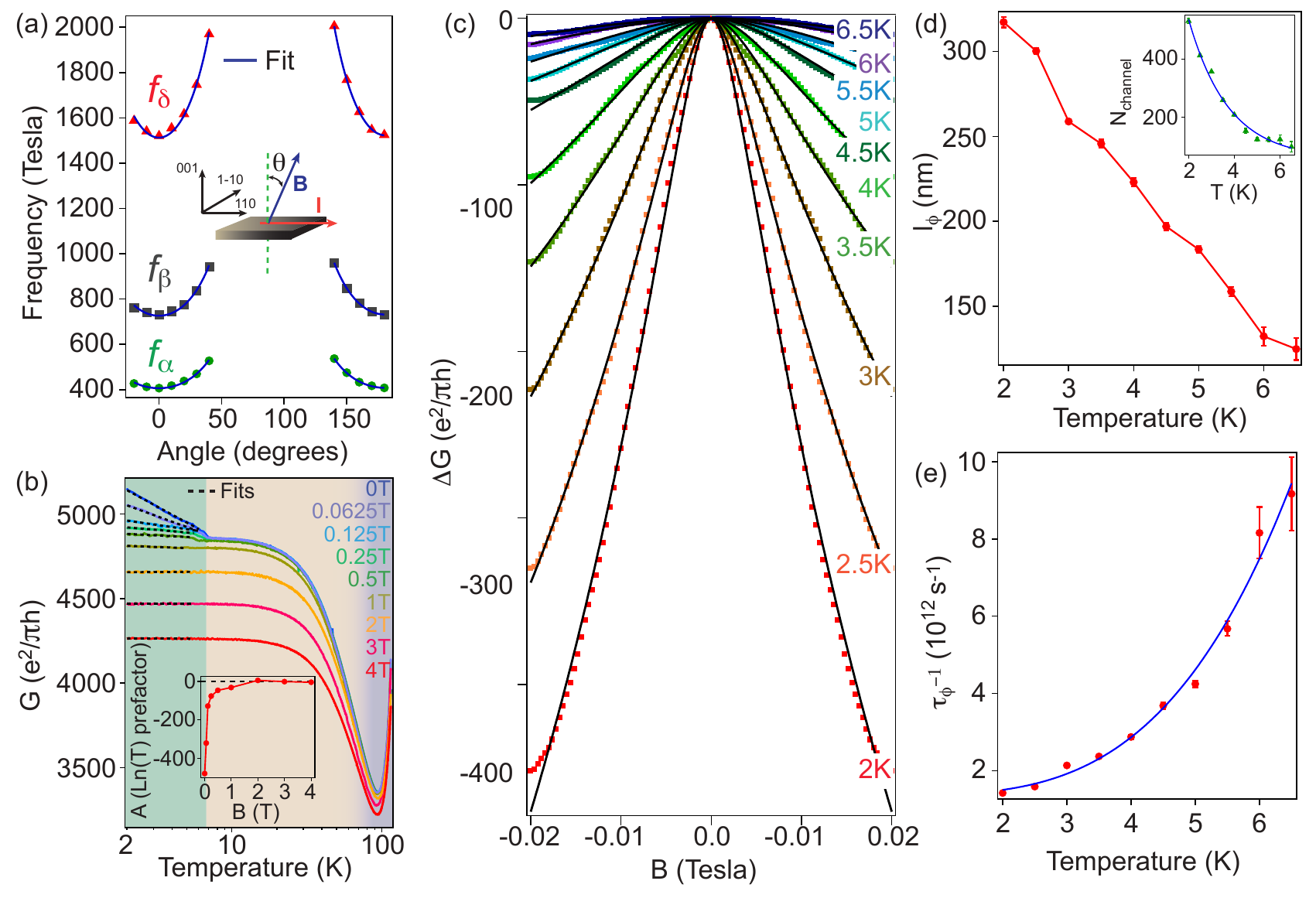}
\caption{Weak antilocalization and quasi-two-dimensional behavior in LuSb/GaSb (001) thin films. (a) Dependence of FFT frequencies on relative angles between the surface normal and the magnetic field direction and corresponding fits to the 2D Fermi surface model as described in the main text. The angle is defined in the inset. (b) Temperature dependence of conductance at different out-of-plane magnetic fields. Background colors indicate transport regimes as described in Fig. 2(a). Fits to quantum corrections to the conductance are shown in dotted black lines. Inset shows the evolution of A, the coefficient of the $ln$(T) term used in the fits, with magnetic field. (c) Evolution of WAL with temperature. Corresponding fits to 2D HLN theory are overlaid as black solid lines. (d) Extracted phase coherence lengths from the fits in (c). Inset shows decrease in the number of channels with increase in temperature as obtained from the same fits in (c). Blue solid line is a fit showing exponential decay in the number of channels with increasing temperature. (e) Inelastic scattering rate as a function of temperature. Blue line is a fit showing T$^{n}$ dependence ($n = 3.47 \pm 0.38$) of the inelastic scattering rate.}
\label{fig:WAL}
\end{figure*}

At low magnetic fields under strong spin-orbit coupling HLN theory predicts quantum correction to conductance under perpendicular magnetic field as\cite{Hikami:wh}
\begin{equation}
  \Delta G_{\perp,WAL}(B) = \alpha N_{channel}\frac{e^{2}}{\pi h}[\Psi(\frac{1}{2} + \frac{B_{\phi}}{B_{\perp}}) - ln(\frac{B_{\phi}}{B_{\perp}})]
 \end{equation}
where $N_{channel}$ is the number of parallel 2D channels, $\Psi$ is the digamma function, $B_{\phi} = \frac{\hbar}{4el_{\phi}^{2}}$ is the characteristic magnetic field corresponding to the phase coherence length $l_{\phi}$, and $\alpha  = -1/2$ in the limit of strong spin-orbit scattering.
We note that the magnitude of the quantum correction effects in our thin films is relatively large, which would generally be construed as arising from three-dimensional carriers. However, evidence provided so far leads us to consider the electronic states as quasi-two-dimensional indicating the presence of a large number of quasi-2D channels in our thin films due to dimensional confinement. Good fits to the WAL data are achieved using HLN theory, as shown in Fig. \ref{fig:WAL}(c). Phase coherence lengths, shown in Fig. \ref{fig:WAL}(d) are found to be much larger than the film thickness, thus validating our initial assumption of the applicability of the HLN theory for a two-dimensional electron gas. We find a T$^{-n}$ dependence of the dephasing time ($\tau_{\phi}$) with $n = 3.47 \pm 0.38$ down to the lowest measured temperature of 2K that can be ascribed to electron-phonon scattering in a two-dimensional electronic system\cite{Lin:2002dt}. We estimate phonon velocity $v_{ph}$  = 1.69 km/s and the characteristic phonon wavelength $\lambda_{ph}$ $\approx$ 40.5nm at 2K\cite{suppl}, which is greater than the film thickness. Therefore, at the lowest measured temperatures phonons in our thin films should be considered as quasi-two-dimensional. The number of independent two-dimensional channels ($N_{channel}$) estimated from the fits decreases exponentially with increasing temperature plausibly due to enhanced inter subband scattering at higher temperatures.\\

\begin{figure}
\includegraphics[width=0.9\columnwidth]{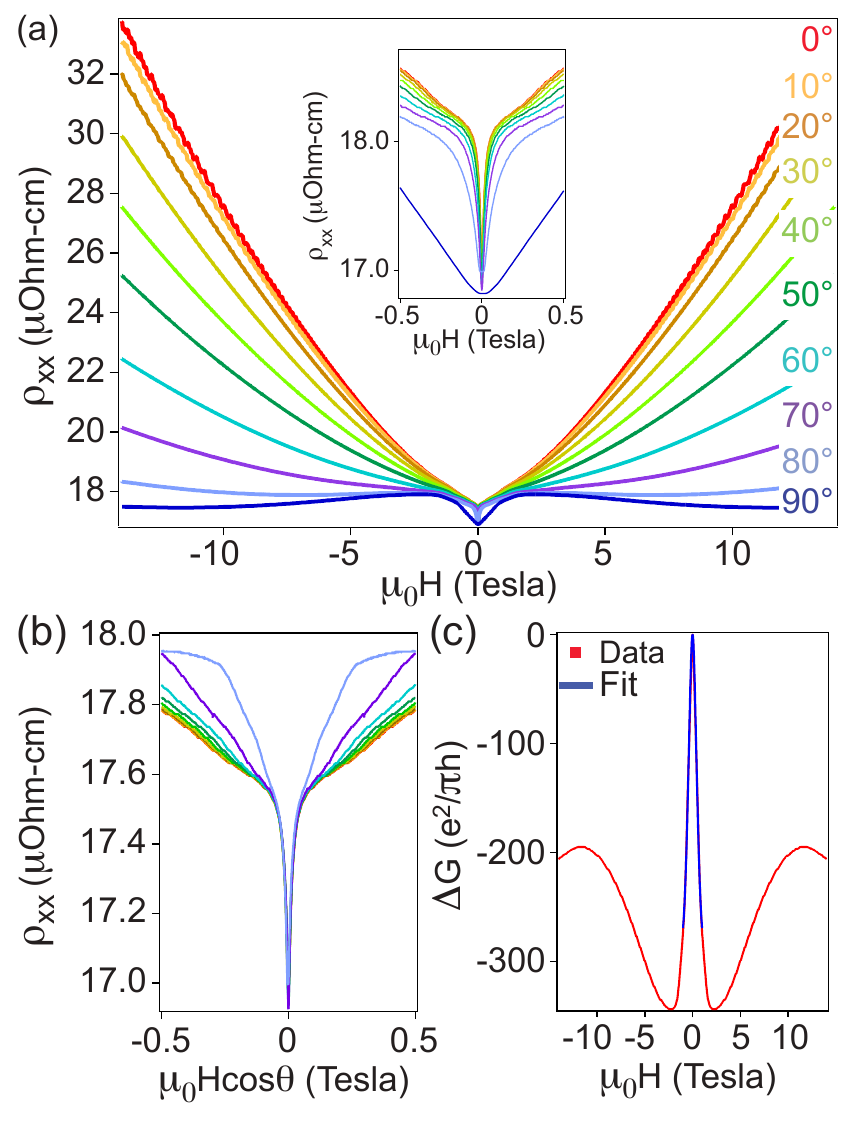}
\caption{Angular dependence of weak antilocalization in LuSb/GaSb (001) thin films. (a) Longitudinal resistance as a function of angle between the surface normal and the magnetic field vector. The angle is defined in the inset of Fig. 4(a). A zoomed-in image between $\pm$0.5 T, showing positive magnetoresistance due to WAL, is shown in the inset. (b) Same data in (a) plotted as a function of the normal component of the applied magnetic field showing that the WAL scales with the normal component of the magnetic field vector between $\pm$0.05 T (c) Parallel field magnetoconductance. Fits to Eq. (2), as described in the text, to the low field magnetoconductance between $\pm$1 T is shown in blue.}
\label{fig:WALtilt}
\end{figure}

 The quantum corrections appearing at low temperatures in our magnetotransport data are found to be sensitive to the normal component of the magnetic field vector, as shown in Figs. \ref{fig:WALtilt}(a) and (b), further underscoring the quasi-2D nature of the electronic states in our thin films. We provide one final piece of evidence for the quasi-two-dimensional nature by examining WAL effects that appear when the magnetic field vector is in the film plane. For an ideal 2D system no WAL induced magnetoresistance is expected in this measurement configuration. However, nonzero electron diffusion in the out-of-plane film direction always results in a finite WAL effect on application of in-plane magnetic field. 2D WAL corrections in such a configuration in the strong spin-orbit regime are given by\cite{AA:1981,DK:1984,Beenakker:1988tf}
\begin{widetext}
\begin{equation}
\Delta G_{\parallel,WAL}(B) = N_{channel}\frac{e^{2}}{\pi h}[\frac{3}{2} ln(1 + \beta\frac{B_{\parallel}^{2}}{B_{d}B_{2}})- \frac{1}{2}ln(1+ \beta \frac{B_{\parallel}^{2}}{B_{d}B_{3}})]
\end{equation}
\end{widetext}

where $B_{d}$ = $\frac{4\hbar}{et^{2}}$ ($t$ is the film thickness) and is equal to 13.2 T in our case, $B_{2}$ = $B_{\phi}$ + $\frac{4}{3}B_{SO}$, $B_{3}$ = $B_{\phi}$. $B_{SO} = \frac{\hbar}{4el_{SO}^{2}}$ is the characteristic magnetic field corresponding to the spin-orbit scattering length $l_{SO}$. We have ignored spin-flip scattering in our LuSb thin films, which is nonmagnetic. The above equation is valid for magnetic field strengths less than 1.65 T beyond which the characteristic magnetic length, given by $l_{B} = \sqrt{\frac{\hbar}{2eB}}$, exceeds the film thickness and transport in our thin films can no longer be considered two dimensional. Our data under parallel magnetic field is described very well by the 2D theory, as shown by the fit in Fig. \ref{fig:WALtilt}(c), with the same $l_{\phi}$ and $N_{channel}$ values as obtained from perpendicular field magnetoresistance, with B$_{SO}$ and $\beta$ as the free parameters. We obtained a spin orbit scattering length of 90.7 nm at 2K, which is much smaller than the phase coherence length of 317.4 nm and a $\beta$ value of 0.097. $\beta$ is expected to be equal to $\frac{1}{3}$, when $t$ $\gg$ $l_{e}$\cite{AA:1981} and is $\frac{1}{16}$($\frac{t}{l_{e}}$) in the opposite limit\cite{DK:1984}. Estimated elastic scattering length ($l_{e}$) in our thin film is greater than that of the film thickness ($t$), which should place us in the Dugaev-Khimel’nitskii limit\cite{DK:1984}. However, our estimated $\beta$ value suggests an intermediate regime scenario\cite{Beenakker:1988tf}, where $\frac{1}{16}$($\frac{t}{l_{e}}$) $<$ $\beta$ $<$ $\frac{1}{3}$, which is attributed to additional contribution from inter subband scattering due to the presence of multiple parallel 2D channels in our thin films. Magnetoconductance in our thin films becomes positive at intermediate field values beyond 2.3 T, reminiscent of the observed parallel field magnetoconductance for 2D channels in III-V semiconductors, suggestive of reduced inter sub-band scattering at higher fields\cite{Englert:1983}. It becomes negative again at a stronger field of $\approx$ 11.4 T when the thin film is firmly in the 3D limit plausibly due to dominant contribution of classical magnetoresistance at high field values.

\section{Methods}
\subsubsection{Thin film growth}
Thin films were grown by molecular-beam epitaxy (MBE) in a MOD Gen II growth chamber. A 5-nm-thick GaSb buffer layer was grown on low n-type doped GaSb (001) substrates (that freezes out at low temperatures, see Ref. \cite{suppl}) at 450$^{\circ}$C under Sb$_{4}$ overpressure after desorption of the native oxide using atomic hydrogen. This is followed by co-evaporation of Lu and Sb from calibrated effusion cells with the substrate temperature at 285$^{\circ}$C - 380$^{\circ}$C and Lu to Sb flux ratio ranging between 1:1 and 1:4. Samples grown at lower Lu:Sb flux ratio and/or lower substrate temperature resulted in films that did not show weak antilocalization effects. Atomic fluxes of Lu and Sb are calibrated by Rutherford backscattering spectrometry (RBS) measurements of the elemental areal density of calibration samples on Si. These measurements were used to calibrate \emph{in situ} beam flux measurements using an ion gauge. Sample surfaces were protected with a 10-nm-thick AlOx layer using e-beam evaporation before taking them out of the UHV chamber. For ARPES measurements conductive n-type Te doped GaSb (001) substrates were used. 
\subsubsection{Low-temperature transport}
Transport measurements were performed on a fabricated hall-bar device using standard a.c. lock-in technique  at low temperatures with the current flowing along [110] crystallographic direction where parallel conduction from the substrate and the buffer layers can be neglected [see Fig. \ref{fig:QO}(a) and Ref. \cite{suppl}). Hall bars were fabricated using standard optical lithography, followed by an ion milling procedure using argon ions. Contacts were made using 50 $\mu$m gold wire bonded onto gold pads. Low temperature measurements were carried out in a Quantum Design PPMS with base temperature of 2K and maximum magnetic field of 14T. 
\subsubsection{ARPES}
Samples were transferred in a custom-built vacuum suitcase from the growth chamber at Santa Barbara to the ARPES endstation 10.0.1.2 at the Advanced Light Source in Berkeley. Pressure inside the vacuum suitcase was better than 1$\times$10$^{-10}$ Torr. Tunable synchrotron light in the 20-80 eV range was used for photoemission measurement with a Scienta R4000 analyzer. The base pressure of the analysis chamber was better than 5$\times$10$^{-11}$ Torr.
\subsubsection{HAADF-STEM}
High-angle annular dark field scanning transmission electron microscopy (HAADF-STEM) was used for imaging the cross section of the epitaxial layer. The cross-sectional lamellas for STEM were prepared using a FEI Helios Dual-beam Nanolab 650 focused gallium ion beam (FIB).  FIB etching steps down to 2KeV were used to polish down the lamella to approximately 50 nm in thickness.
\subsubsection{Computational approach}
The calculations were based on density functional theory (DFT)\cite{hohenberg1964inhomogeneous,kohn1965self} and the
hybrid functional of Heyd, Scuseria, and Ernzerhof
(HSE06)\cite{Heyd:2003eg,HSE} as implemented in the VASP code\cite{kresse1993ab,kresse1994ab}.
The interaction between the valence electrons and the ionic cores was described using projector augmented-wave (PAW) potentials \cite{blochl1994projector,kresse1999ultrasoft}. 
The PAW potential for Sb has five valence electrons with 5s\textsuperscript{2}5p\textsuperscript{3} configuration, whereas for Lu there are nine valence electrons, i.e., 5p\textsuperscript{6}6s\textsuperscript{2}5d\textsuperscript{1} configuration. Test calculations including the localized Lu 4$f$ orbitals in the valence showed a dispersionless fully occupied 4$f$ bands at $\sim8$ eV below the Fermi level, and change the calculated carrier density by less than 5\%. We used a plane-wave basis set with 300 eV kinetic energy cutoff and 8x8x8 $\Gamma$-centered mesh of k points for integrations over the Brillouin zone of the primitive cell of rock-salt crystal structure with two atoms, one located at (0,0,0) and the other at (0.5,0.5,0.5). SuperCell K-space Extremal Area Finder (SKEAF)\cite{Julian2012} and Wannier90\cite{Mostofi2014} codes were used for the calculation of carrier density and SdH frequencies, whereas the effective mass was calculated at the Fermi level by getting the second derivative.

\section{Conclusion}

In summary, we have demonstrated our ability to synthesize high quality LuSb thin films and controllably introduce defects to access diffusive regime in transport measurements. By employing quantum oscillations, ARPES, and DFT calculations we have thoroughly characterized its electronic structure that establishes LuSb as a compensated semimetal and topologically trivial. Large phase coherence lengths coupled with strong spin-orbit scattering led to the observation of weak antilocalization at low temperatures. Quasi-two-dimensionality of the electronic states, significant reduction of the Debye temperature from its bulk value, and accessibility to the two-dimensional limit of the phonon spectrum offers opportunities to control electron-phonon coupling in epitaxial thin films. Furthermore, recent DFT calculations predict the possibility of a topological phase transition in LuSb and LuBi on application of bi-axial strain\cite{Narimani:2018bc}, which should now be accessible to experimentalists. Our work lays the foundation for further studies of controlled tunability of the electronic properties via epitaxial strain, dimensional confinement  and electrostatic gating in this technologically relevant material system for novel device applications.

\section*{Acknowledgements}
The authors thank S. James Allen for helpful discussions. Synthesis of thin films, development of the UHV suitcase, ARPES experiments, and theoretical work are supported by the US Department of Energy (Contract No. DE-SC0014388). Development of the growth facilities and low temperature magnetotransport measurements are supported by the Office of Naval Research through the Vannevar Bush Faculty Fellowship under the Award No. N00014-15-1-2845. This research used resources of the Advanced Light Source, which is a DOE Office of Science User Facility under contract no. DE-AC02-05CH11231. We acknowledge the use of facilities within the National Science Foundation Materials Research and Engineering Center (DMR 11-21053) at the University of California at Santa Barbara, the LeRoy Eyring Center for Solid State Science at Arizona State University. Density functional theory calculations made use of the Extreme Science and Engineering Discovery Environment, NSF Grant No. ACI-1053575, and high-performance computing and the Information Technologies resources at the University of Delaware. D.R. gratefully acknowledges support from the Leverhulme Trust via an International Academic Fellowship (IAF-2018-039).

\end{document}


\title{Weak antilocalization in quasi-two-dimensional electronic states of epitaxial LuSb thin films : Supplementary Information}

\author{Shouvik Chatterjee}
\email[Authors to whom correspondence should be addressed: ]{schatterjee@ucsb.edu, cjpalm@ucsb.edu}
\affiliation{Department of Electrical $\&$ Computer Engineering, University of California, Santa Barbara, CA 93106, USA}
\author{Shoaib Khalid}
\affiliation{Department of Physics and Astronomy, University of Delaware, Newark, DE 19716, USA}
\affiliation{Department of Materials Science and Engineering, University of Delaware, Newark, DE 19716, USA}
\author{Hadass S. Inbar}
\affiliation{Materials Department, University of California, Santa Barbara, CA 93106, USA}
\author{Aranya Goswami}
\affiliation{Department of Electrical $\&$ Computer Engineering, University of California, Santa Barbara, CA 93106, USA}
\author{Felipe Crasto de Lima}
\affiliation{Department of Materials Science and Engineering, University of Delaware, Newark, DE 19716, USA}
\author{Abhishek Sharan}
\affiliation{Department of Physics and Astronomy, University of Delaware, Newark, DE 19716, USA}
\affiliation{Department of Materials Science and Engineering, University of Delaware, Newark, DE 19716, USA}
\author{Fernando P. Sabino}
\affiliation{Department of Materials Science and Engineering, University of Delaware, Newark, DE 19716, USA}
\author{Tobias L. Brown-Heft}
\affiliation{Materials Department, University of California, Santa Barbara, CA 93106, USA}
\author{Yu-Hao Chang}
\affiliation{Materials Department, University of California, Santa Barbara, CA 93106, USA}
\author{Alexei V. Fedorov}
\affiliation{Advanced Light Source, Lawrence Berkeley National Laboratory, Berkeley, California 94720, USA}
\author{Dan Read}
\affiliation{School of Physics and Astronomy, Cardiff University, Cardiff CF24 3AA, UK}
\author{Anderson Janotti}
\affiliation{Department of Materials Science and Engineering, University of Delaware, Newark, DE 19716, USA}
\author{Christopher J. Palmstr\o m}
\email[Authors to whom correspondence should be addressed: ]{schatterjee@ucsb.edu, cjpalm@ucsb.edu}
\affiliation{Department of Electrical $\&$ Computer Engineering, University of California, Santa Barbara, CA 93106, USA}
\affiliation{Materials Department, University of California, Santa Barbara, CA 93106, USA}
\date{\today}
\maketitle

\section{Dependence of transport properties on growth conditions: }
Growth temperature and Sb to Lu beam flux ratio has a strong influence on the synthesis of epitaxial LuSb thin films. By raising the growth temperature or by reducing Sb to Lu beam flux ratio from their optimal values it is possible to controllably introduce defects in the synthesized thin films such that signatures of weak antilocalization can be observed at low temperatures. In Fig. S1 we show Scanning Electron Microscopy (SEM) images of LuSb thin films synthesized under different growth conditions. Sample A is synthesized at a substrate temperature of 380$^{\circ}$C, whereas Sample B is synthesized at 285$^{\circ}$C  with Sb to Lu flux ratio $\approx$ 1. In contrast to Sample B, SEM image of Sample A reveals pronounced extended defects and also exhibits signatures of weak antilocalization at low temperatures. 
Fast Fourier Transform (FFT) of the Shubnikov de-Haas (SdH) oscillations for both the films is shown in Fig. S1(e). Fermi wave vectors ($k_{F}$) for the Fermi pockets is derived from the corresponding characteristic frequency ($f_{FFT}$) using the Onsager relation, $A = \frac{2\pi^{2}f_{FFT}}{\Phi_{0}}$, where \textit{A} is the projected area of the Fermi surface normal to the magnetic field vector and $\Phi_{0}$ is the magnetic flux quantum. For the $\beta$ and the $\delta$ pockets a spherical cross-section is assumed, while for the elliptical $\alpha$ pocket Fermi wave vectors along the semiminor($k_{F,a}$) and semimajor axes($k_{F,b})$ is derived using $A_{\alpha} = \pi k_{F,a}^{2}$ and $A_{\alpha_{I}} = \pi k_{F,a}k_{F,b}$. Carrier concentration is derived using n$_{3D}$ = $\frac{V_{F}}{4\pi^{3}}$, where $V_{F}$ is the volume of the Fermi surface. For the $\beta$ and the $\delta$ hole pockets $V_{F}$ = $\frac{4}{3}\pi k_{F}^{3}$, whereas for the elliptical $\alpha$ pocket $V_{F}$ = $\frac{4}{3}\pi k^{2}_{F,a}k_{F,b}$. The level of carrier compensation is determined from the ratio of the total electron (n$_{e}$) and holelike (n$_{h}$) carriers considering three electron pockets ($\alpha$) and two holelike pockets ($\beta$ and $\delta$) 

\begin{equation}
\frac{n_{h}}{n_{e}} = \frac{n_{\beta} + n_{\delta}}{3n_{\alpha}}
\end{equation}

which is found to be 1.06 and 1.05 for the samples A and B, respectively. Though, carrier concentrations remain nearly identical in the two samples, changing growth temperatures has a substantial effect on the relative mobility values of the carriers resulting in a change from electronlike (sample B) to holelike behavior (sample A) in the Hall voltage as shown in Fig. S1(c) and (d).
\newpage
\begin{center}
\begin{table*}
\caption{Fermi Surface in Sample A and Sample B}
\begin{ruledtabular}
\begin{tabular}{c@{\quad}ccc@{\quad}ccc}

  & \multicolumn{2}{c}{Sample A} & \multicolumn{2}{c}{Sample B} \\
 FS\footnote{FS denotes Fermi Surface} & $f_{SdH}$ (Tesla)  & n$_{SdH}$ (cm$^{-3}$)  & $f_{SdH}$ (Tesla) & n$_{SdH}$ (cm$^{-3}$)\\
  \hline
  $\alpha$ & 415($\alpha$), 1245($\alpha_{I}$)  & 1.435$\times$10$^{20}$ & 415($\alpha$), 1245($\alpha_{I}$) &  1.435$\times$10$^{20}$ \\
  $\beta$ & 743  & 1.15$\times$10$^{20}$ & 740 &  1.14$\times$10$^{20}$ \\
  $\delta$ & 1544  & 3.43$\times$10$^{20}$ & 1535 &  3.40$\times$10$^{20}$ \\

\end{tabular}
\end{ruledtabular}
\end{table*}
\end{center}

\section{Estimation of transport parameters:}
An anisotropic multi-band model that explicitly takes into account ellipticity of the $\alpha$ pocket is used to simultaneously fit the low field magnetoresistance and Hall data to estimate the mobility values of the individual carriers. Ratio of the mobility values along the semiminor ($\mu_{a}$) and semimajor ($\mu_{b}$) axes of the $\alpha$ pocket is given by $\xi = \frac{\mu_{a}}{\mu_{b}} = (\frac{k_{F,b}}{k_{F,a}})^{2}$ = 9.005. For magnetic field along [001]  and current along [110] crystallographic directions, taking into account the anisotropic mobility of the $\alpha$ pocket, we obtain

\begin{equation}
\begin{split}
& \sigma^{\alpha_{x}}_{x^{'}x^{'}} = \sigma^{\alpha_{y}}_{x^{'}x^{'}} =  \frac{en_{\alpha}\mu_{a}\sqrt{\frac{(1+\xi^{2})}{2}}}{1+\frac{(1+\xi^{2})}{2}(\mu_{a}B)^{2}} \\
& \sigma^{\alpha_{z}}_{x^{'}x^{'}} = \frac{en_{\alpha}\xi\mu_{a}}{1+(\xi\mu_{a}B)^{2}}\\
&\sigma^{\alpha_{x}}_{x^{'}y^{'}} = \sigma^{\alpha_{y}}_{x^{'}y^{'}} =  \frac{en_{\alpha}\mu^{2}_{a}(1+\xi^{2})B}{1+(1+\xi^{2})(\mu_{a}B)^{2}}\\
& \sigma^{\alpha_{z}}_{x^{'}y^{'}} = \frac{en_{\alpha}\xi^{2}\mu^{2}_{a}B}{1+(\xi\mu_{a}B)^{2}}\\
\end{split}
\end{equation}
where unit vectors $\hat{\mathbf{x}}$, $\hat{\mathbf{y}}$, $\hat{\mathbf{z}}$, $\hat{\mathbf{x^{'}}}$, and $\hat{\mathbf{y^{'}}}$ are along [100], [010], [001], [110] and [1$\bar{1}$0] crystallographic directions, respectively.

For the hole pockets ($\beta$, $\delta$) mobility is assumed to be isotropic. Therefore,
\begin{equation}
\begin{split}
& \sigma^{i}_{x^{'}x^{'}} = \frac{en_{i}\mu^{i}}{1+(\mu^{i}B)^{2}} \\
& \sigma^{i}_{x^{'}y^{'}} = \frac{en_{i}\mu^{i}}{1+(\mu^{i}B)^{2}},    i = \beta, \delta
\end{split}
\end{equation}

Finally, resistivity is obtained as

\begin{equation}
    \begin{split}
       & \rho_{x^{'}x^{'}} = \frac{\Sigma_{i}\sigma^{i}_{x^{'}x^{'}}}{(\Sigma_{i}\sigma^{i}_{x^{'}x^{'}})^{2} + (\Sigma_{i}\sigma^{i}_{x^{'}y^{'}})^{2} }\\
       & \rho_{x^{'}y^{'}} = \frac{-\Sigma_{i}\sigma^{i}_{x^{'}y^{'}}}{(\Sigma_{i}\sigma^{i}_{x^{'}x^{'}})^{2} + (\Sigma_{i}\sigma^{i}_{x^{'}y^{'}})^{2} }, i = \alpha^{x}, \alpha^{y}, \alpha^{z}, \beta, \delta
    \end{split}
\end{equation}

Fits to the data at 8K, both for sample A and sample B is shown in Fig. S2. Obtained mobility values are listed in Table S2. Relevant transport parameters viz. elastic scattering length ($l_{e}$), Fermi velocity ($v_{F}$), Diffusion coefficient (\textit{D}), and characteristic magnetic field ($B_{e}$) listed in Table 2 of the main text are calculated as follows:
\begin{equation}
    \begin{split}
     & v_{F} = \frac{\hbar k_{F}}{m^{*}}\\
     & D = v_{F}\tau \\
     & B_{e} = \frac{\hbar}{4eD\tau_{e}}\\
     & l_{e} = \sqrt{D\tau_{e}}\\
    \end{split}
\end{equation}
where $m^{*}$ is the effective mass, $k_{B}$ is the Boltzmann constant, $T$ is the temperature in Kelvin, and $\tau_{e}$ is the elastic scattering time given as $\tau_{e} = m^{*}\mu/e$
\newpage
\newpage
\begin{center}
\begin{table*}
\caption{Mobility values in Sample A and Sample B}
\begin{ruledtabular}
\begin{tabular}{c@{\quad}ccc@{\quad}ccc}
	& \multicolumn{2}{c}{$\mu$(cm$^{2}$ / Vs)} \\

  Fermi Surface & Sample A & Sample B \\
  \hline
  $\alpha$\footnote{Average mobility for the $\alpha$ pocket is taken as the geometric mean of the mobility values along the semimajor($\hat{\mathbf{b}}$) and two semiminor axes($\hat{\mathbf{a}}$) of the ellipsoid $\mu^{\alpha}_{avg} = \mu^{1/3}_{a}\mu^{2/3}_{b}$} & 5.42$\times$10$^{2}$   & 1.739$\times$10$^{3}$\\
  $\beta$ & 4.078$\times$10$^{3}$   & 2.411$\times$10$^{3}$\\
  $\delta$ & 2.036$\times$10$^{3}$   & 1.359$\times$10$^{3}$\\
 
\end{tabular}
\end{ruledtabular}
\end{table*}
\end{center}

\section{Estimation of phonon wavelength: }

For temperatures much less than the Debye temperature ($\Theta_{D}$) characteristic phonon wavelength $\lambda_{phonon}$ is given by
\begin{equation}
   \lambda_{phonon} = \frac{hv_{s}}{k_{B}T} 
\end{equation}
where $v_{s}$, $k_{B}$ and $T$ are the velocity of sound, the Boltzmann constant and temperature in Kelvin, respectively. $v_{s}$ can be obtained from the Debye temperature as it is related to the minimum possible wavelength of phonon ($\lambda_{min} = \frac{hv_{s}}{k_{B}\Theta_{D}}$), which should approximately be equal to the distance between the nearest neighbor atoms in the lattice. In LuSb, $\lambda_{min}$ $\approx$ 0.3 nm and from Bloch-Gr{\"u}neissen analysis, as shown in Fig. 2(a) of the main text, we obtained $\Theta_{D}$ = 267.8 K. From the above considerations we obtain the phonon velocity in LuSb $v_{s}$ = 1.69 km/s that gives us a phonon wavelength $\lambda_{phonon}$ $\approx$ 40.5 nm at 2K.

\section{Critical magnetic field value for Shubnikov-deHaas oscillations in tilted fields:}

Finite thickness of thin films along the growth direction limits the observation of Shubnikov-deHaas (SdH) oscillations for three-dimensional carriers for magnetic field values lower than a critical magnetic field ($B_{c}$) under tilted field geometries.  This is due to the collision of the orbiting electrons with the thin film surface that precludes the completion of a cyclotron orbit. Assuming a classical trajectory of the charge carriers, radius ($r$) of the orbit in the real space can be related to the Fermi wave vector ($k_{F}$) as
\begin{equation}
r = \frac{\hbar k_{F}}{eB}
\end{equation}
Therefore, the critical field ($B_{c}$) at a particular tilt angle $\theta$ between the magnetic field vector and surface normal can be estimated as
\begin{equation}
B_{c} = \frac{2\hbar k_{F}sin\theta}{et}
\end{equation}
where $t$ is the film thickness. The critical field value thus obtained sets the lower bound for $B_{c}$ as only condition imposed here is that the cyclotron orbit fits inside the thin film. An analysis of the onset of SdH oscillations at different tilt angles for the carriers in epitaxial LuSb thin films is shown in Fig. S3 that clearly establishes the two-dimensional nature of the electronic carriers in our epitaxial LuSb thin films.

\begin{figure}
\includegraphics[width=0.7\textwidth]{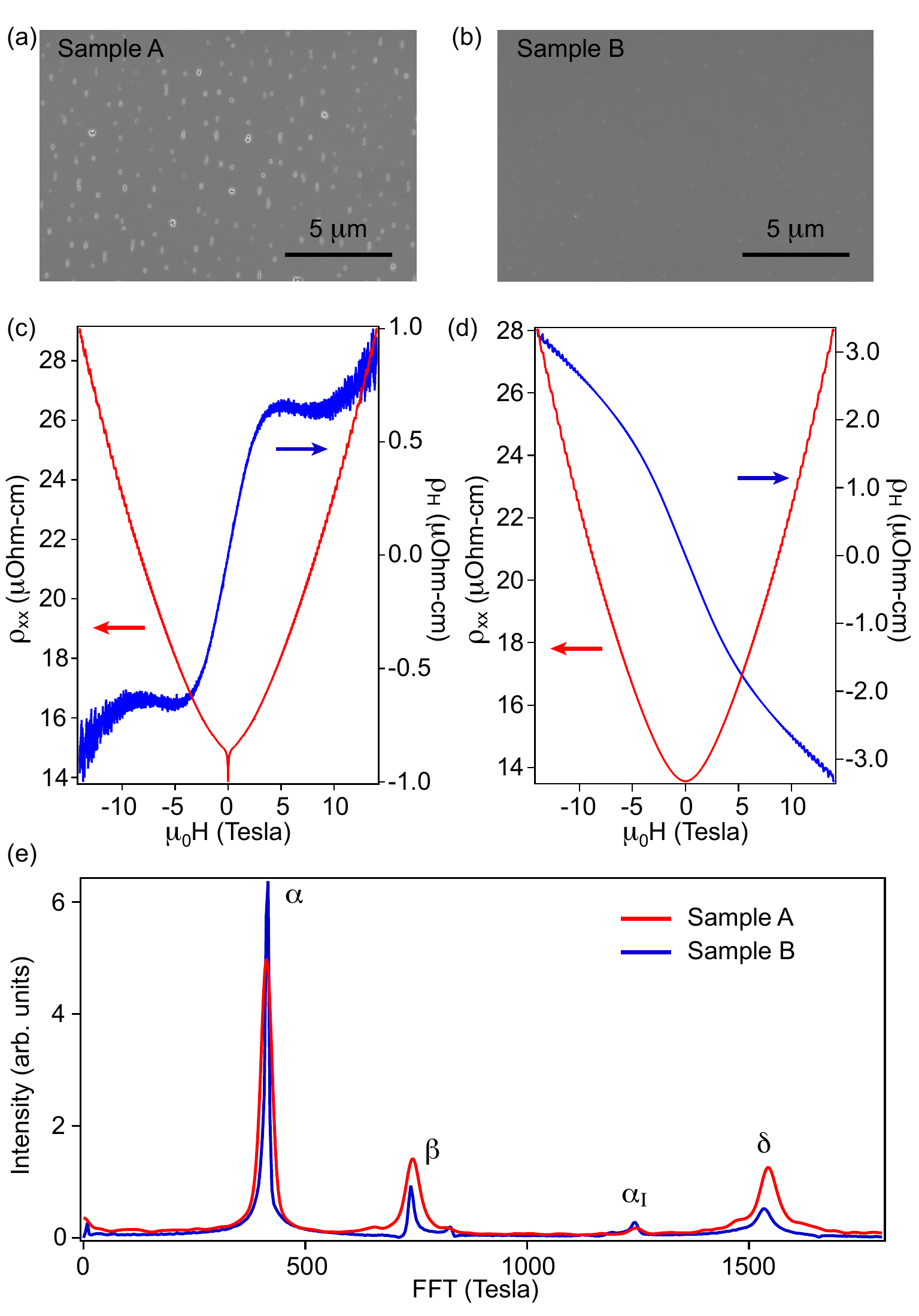}
\caption{\textbf{Introduction of defects in LuSb thin films via growth parameter tuning.} SEM image of samples grown at substrate temperatures of a) 380$^{\circ}$C (sample A) and b) 285$^{\circ}$C (sample B). Sb to Lu beam flux ratio was kept $\approx$ 1 for both the samples. Magnetoresistance and Hall data for c) sample A and d) sample B. e) Fast Fourier Transform (FFT) of the Shubnikov-deHaas (SdH) oscillations from the magnetoresistance data shown in (c) and (d) showing characteristic frequencies corresponding to Fermi surface pockets.}
\label{fig:Defects}
\end{figure}

\begin{figure}
\includegraphics[width=1\textwidth]{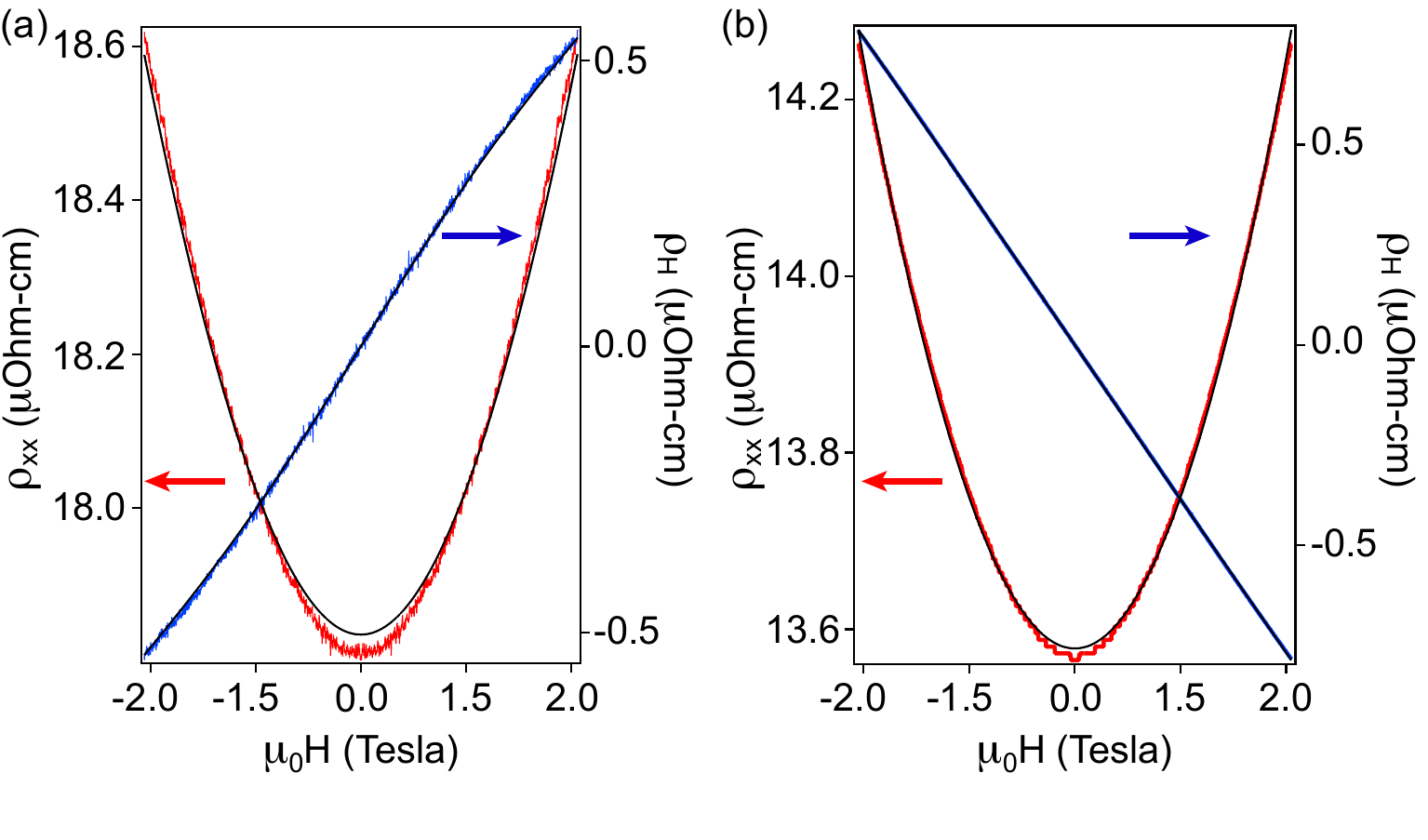}
\caption{\textbf{Fits to transport data using anisotropic multi-band model.} Results for a) sample A and b) sample B. Fits are shown in black lines.}
\label{fig:AnisotropicModel}
\end{figure}

\begin{figure}
\includegraphics[width=0.5\textwidth]{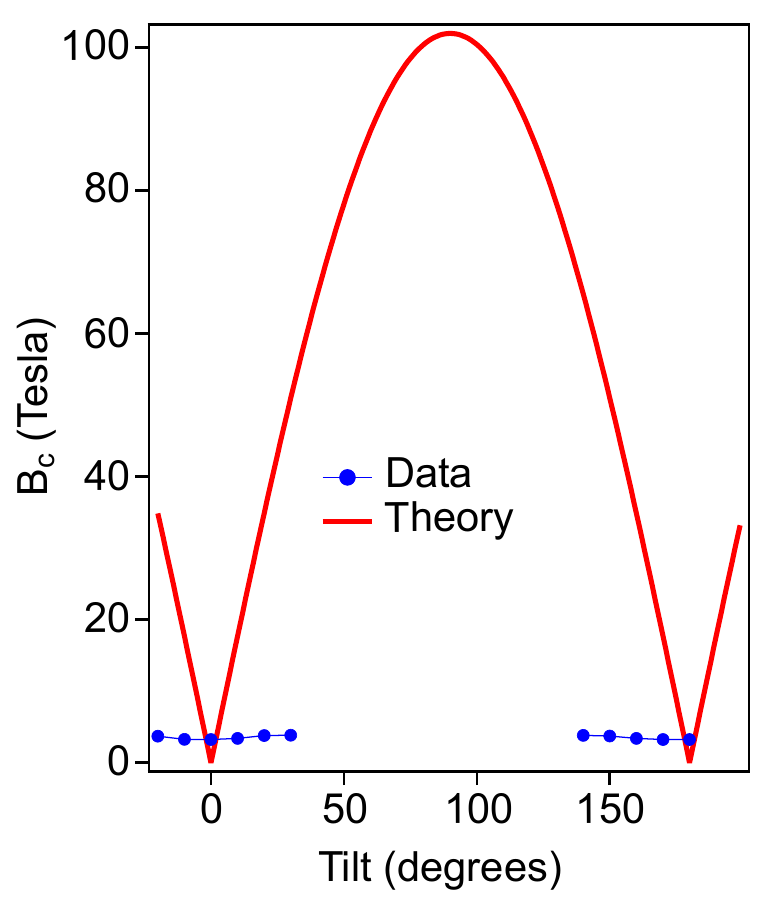}
\caption{\textbf{Critical Field for Shubnikov-deHaas (SdH) oscillations in LuSb thin films.} Estimated critical field ($B_{c}$) required for the observation of SdH oscillation in epitaxial LuSb thin films is shown in red line. We have used $k_{F}$ value of 0.11\AA$^{-1}$, smallest value obtained among all the electronic states in the epitaxial LuSb thin films. Therefore, the estimated critical field represents the lower bound. Experimental data for the onset of SdH oscillations is shown in blue.}
\label{fig:CriticalField}
\end{figure}

\begin{figure}
\includegraphics[width=1\textwidth]{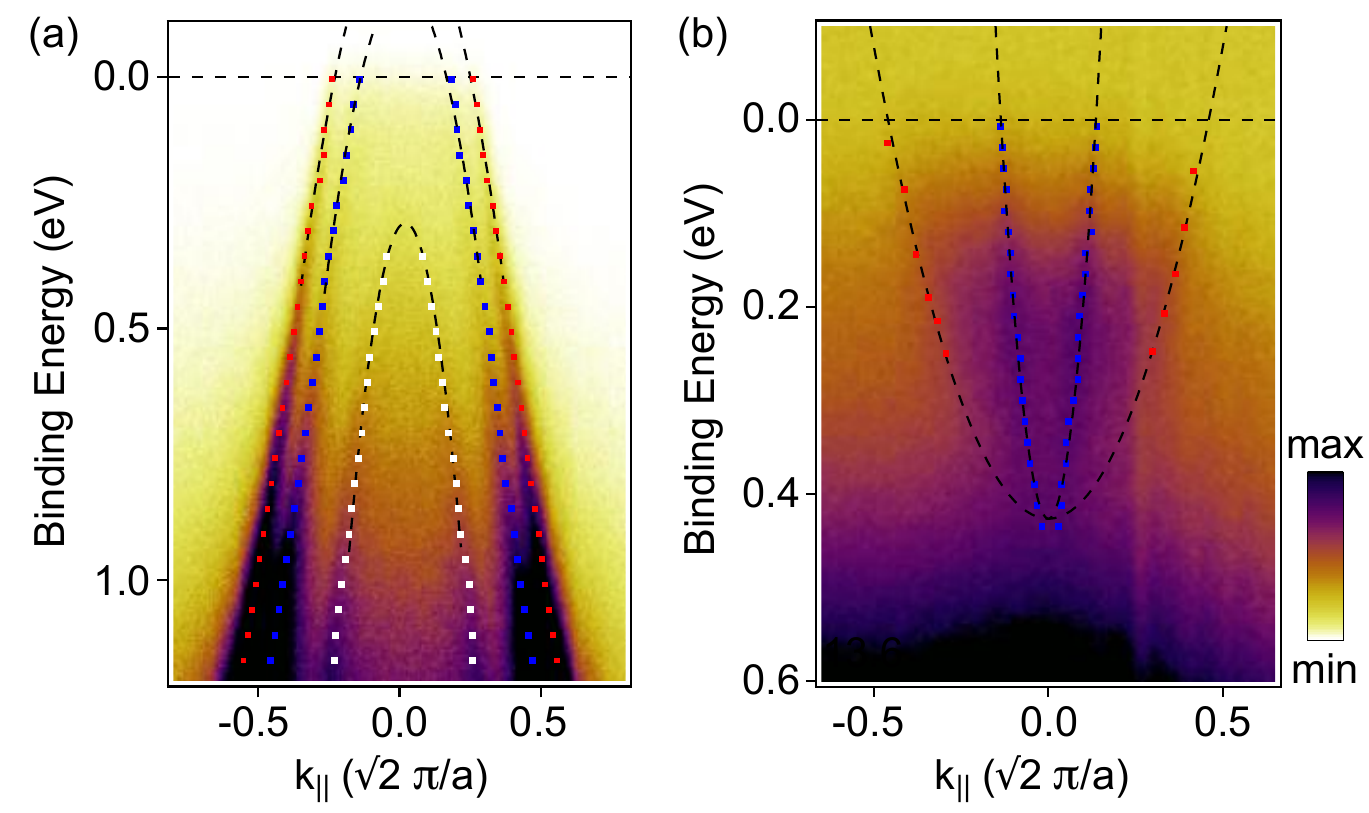}
\caption{\textbf{Parabolic fits to the ARPES data.} Extracted dispersions for the a) hole pockets and b) electron pocket. Extracted dispersions for the $\delta$, $\beta$ and $\gamma$ bands along [110] ($\mathbf{\bar{X}}$-$\mathbf{\bar{\Gamma}}$-$\mathbf{\bar{X}}$) are shown in panel a) in red, blue and white, respectively. Extracted dispersions along the semimajor and semiminor axes of the elliptical $\alpha$ band is shown in panel b) in red and blue, respectively. Parabolic fits are shown in black dotted line.}
\label{fig:ARPESFits}
\end{figure}

\begin{figure}
\includegraphics[width=1\textwidth]{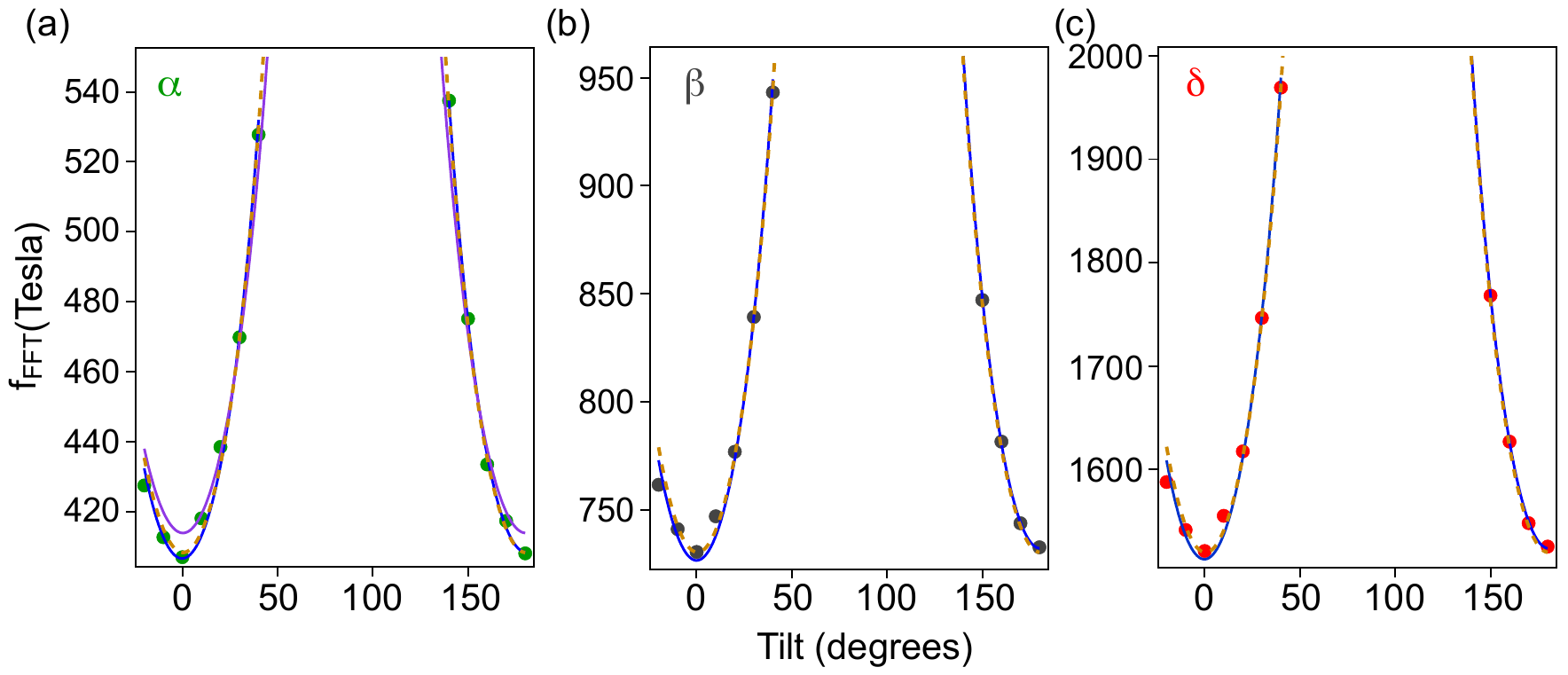}
\caption{\textbf{Evolution of FFT frequencies in LuSb thin films under tilted fields.} Dependence of FFT frequencies on the relative angle ($\theta$) between the magnetic field vector and the surface normal (tilt) for a) $\alpha$ b) $\beta$ and c) $\delta$ bands. Measurement geometry is shown in the inset of Fig. 4(a) of the main text. The dependence is well captured by the two-dimensional Fermi surface model ($f_{\theta} = f_{0} / cos(\theta)$) shown in blue for all the three bands. For the $\alpha$ band, ellipticity of the Fermi surface can result in a similar two-dimensional behavior. But, the experimentally observed ellipticity ($k_{semimajor}/k_{semiminor}$) of 3, as shown in violet in panel a), does not fit well the observed dependence. For an elliptical three-dimensional model to fit the observed dependence we need to invoke unphysical  ellipticity values of 17.5, 12, and 41 for the $\alpha$, $\beta$, and $\delta$ bands, respectively as shown with dotted yellow lines. }
\label{fig:Tilt}
\end{figure}

\begin{figure}
\includegraphics[width=0.7\textwidth]{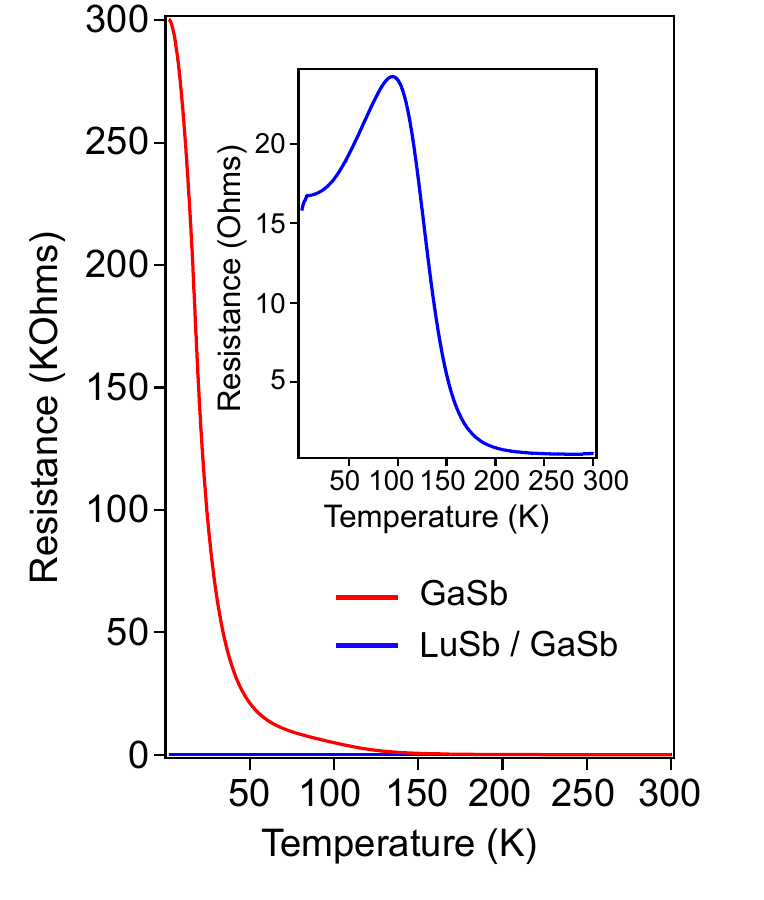}
\caption{\textbf{Comparison of resistance between GaSb and LuSb.} Resistance of a 5 nm thick GaSb buffer layer on an unintentionally doped GaSb substrate (red line) and a 14.2 nm thick LuSb thin film grown on top of the buffer layer (blue). Inset shows the same data as in the main figure, but just for the LuSb thin film highlighting that the resistance in LuSb layers is lower by orders of magnitude in comparison to the GaSb buffer layers / GaSb substrate, particularly at low temperatures. }
\label{fig:Resistance}
\end{figure}

\begin{figure}
\includegraphics[width=0.8\textwidth]{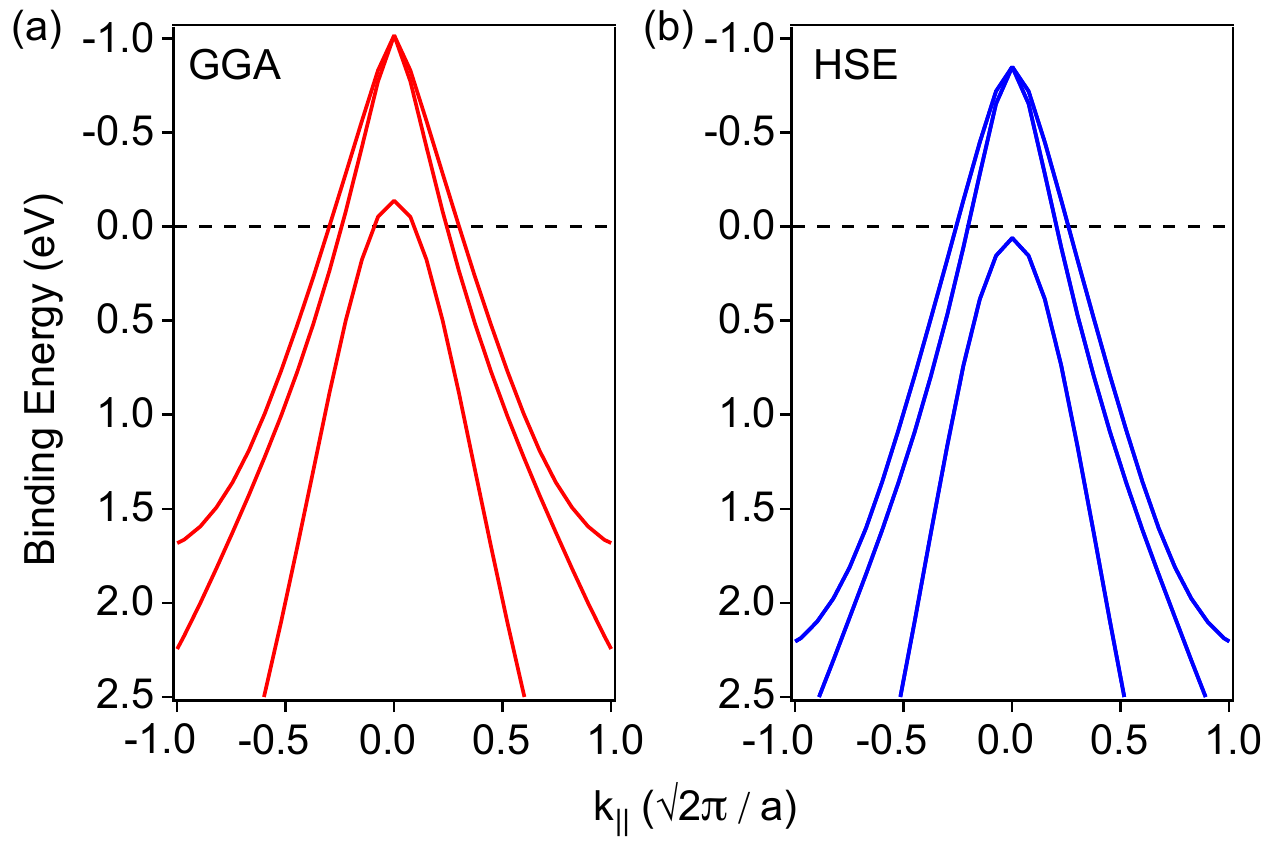}
\caption{\textbf{Comparison between GGA and HSE hybrid functionals.} Band structure calculations of  LuSb along [110] ($\mathbf{\bar{X}}$-$\mathbf{\bar{\Gamma}}$-$\mathbf{\bar{X}}$) crystallographic direction using a) generalized gradient approximation (GGA) and b) Heyd-Scuseria-Ernzerhof (HSE) hybrid functionals. All three hole pockets cross the Fermi level for the calculation that uses GGA functional in contrast to both our experimental findings and predictions from calculations using HSE hybrid functional.}
\label{fig:Resistance}
\end{figure}